\documentclass[letterpaper]{jpconf}
\usepackage{graphicx}
\begin{document}
\setlength{\topmargin}{-1cm}
\setlength{\textwidth}{20cm}
\title{Recent two-photon results at Belle}
\author{H. Nakazawa (Belle Collaboration)}
\address{National Central University (Taiwan)}
\begin{abstract}
 We review recent measurements of pure neutral final state
 production,
 $\gamma\gamma\to\pi^0\pi^0$ and $\eta\pi^0$, 
 and observations of new charmoniumlike resonances,
 $X(3915)$ and $X(4350)$,
 in the two-photon processes
 at the Belle experiment.
\end{abstract}
\section{Introduction}
Two-photon production of exclusive hadronic final states
provides useful information about resonances and pertubative and
nonperturbative QCD.
From theoretical viewpoint, two-photon process is
attractive because of the absence of strong interactions 
in the initial state and the possibility of 
calculating $\gamma\gamma\to q\bar{q}$ amplitudes.
In addition, the quantum numbers of the final state are restricted
to states of charge conjugation $C = +1$ with $J = 1$ forbidden.

We have measured charged pion pair~\cite{jpsj,f0(980), kkpipi},
charged kaon pair~\cite{kkpipi, kk},
neutral kaon pair~\cite{ksks},
proton antiproton pair~\cite{ppbar} and $D$-meson pair~\cite{chic2'} production
in two-photon collisions.
The statistics of these measurements is 2 to 3 orders of magnitude
highter than in the pre-$B$-factory measurements, opening a new
era in studies of two-photon physics.

In this report, we summarize measurements of neutral final state production,
$\gamma\gamma\to\pi^0\pi^0$ and $\eta\pi^0$ and observations of
charmoniumlike resonances $X(3915)$ and $X(4350)$.
\section{Neutral Pair Production}

We use the data samples of 95~fb$^{-1}$~\cite{pi0pi01} and
223~fb$^{-1}$~\cite{pi0pi02} for $\gamma\gamma \to \pi^0\pi^0$
and of  223~pb$^{-1}$ for $\gamma\gamma \to \eta\pi^0$~\cite{etapi0} collected with
the Belle detector~\cite{belle} at the energy-asymmetric
$e^+e^-$ KEKB collider~\cite{kekb}.
Our analysis is based on the ``zero-tag'' mode, where by collecting
small total transverse momentum events,
$|\sum \vec{p}_t|<0.05\,{\rm GeV}/c$, 
the incident photons are guaranteed
to have small virtuality.

\subsection{Light Resonance Study}
\begin{center}
 \begin{figure}
  \begin{minipage}{0.5\hsize}
   \includegraphics[width=6cm]{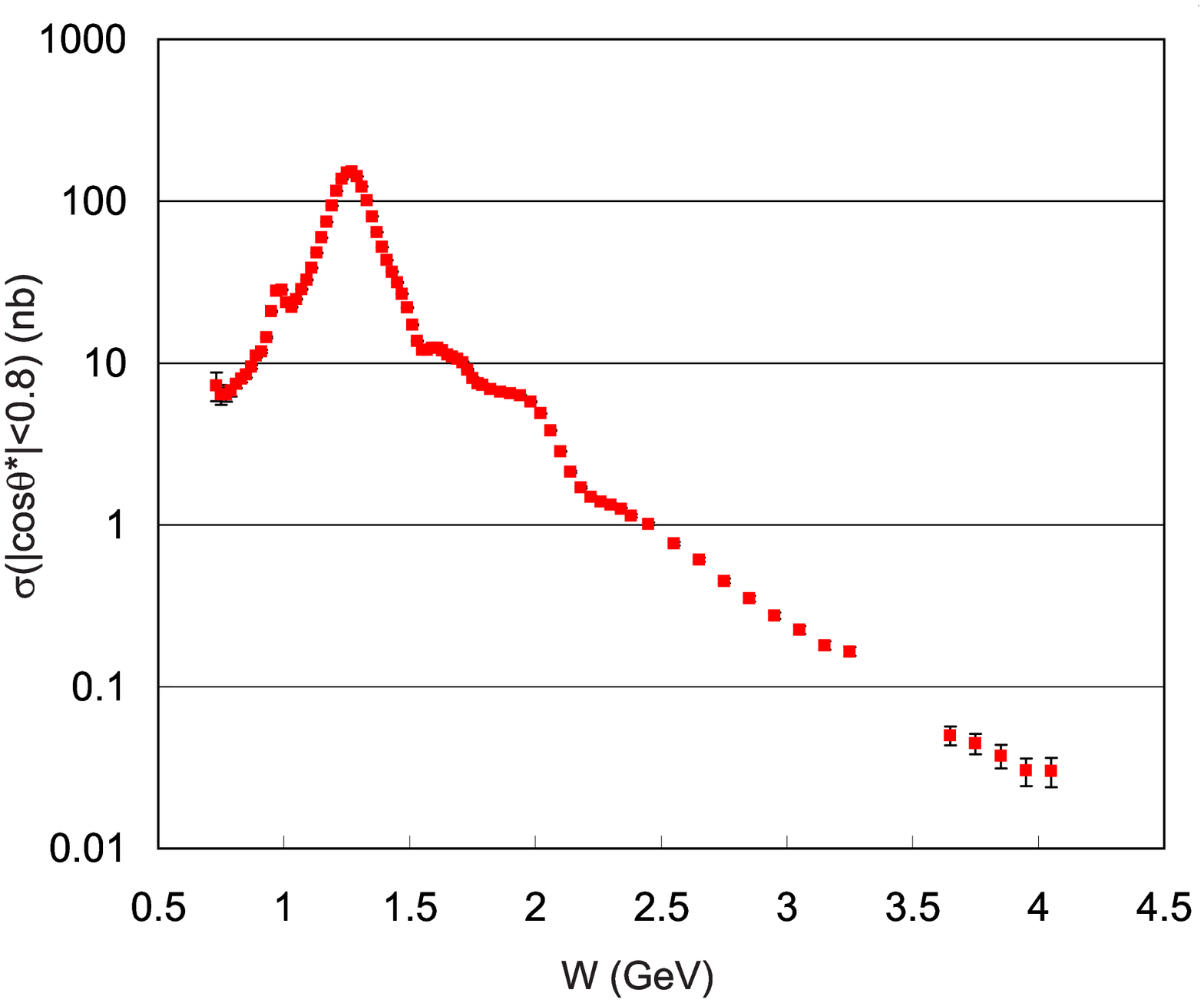}
   \includegraphics[width=6cm]{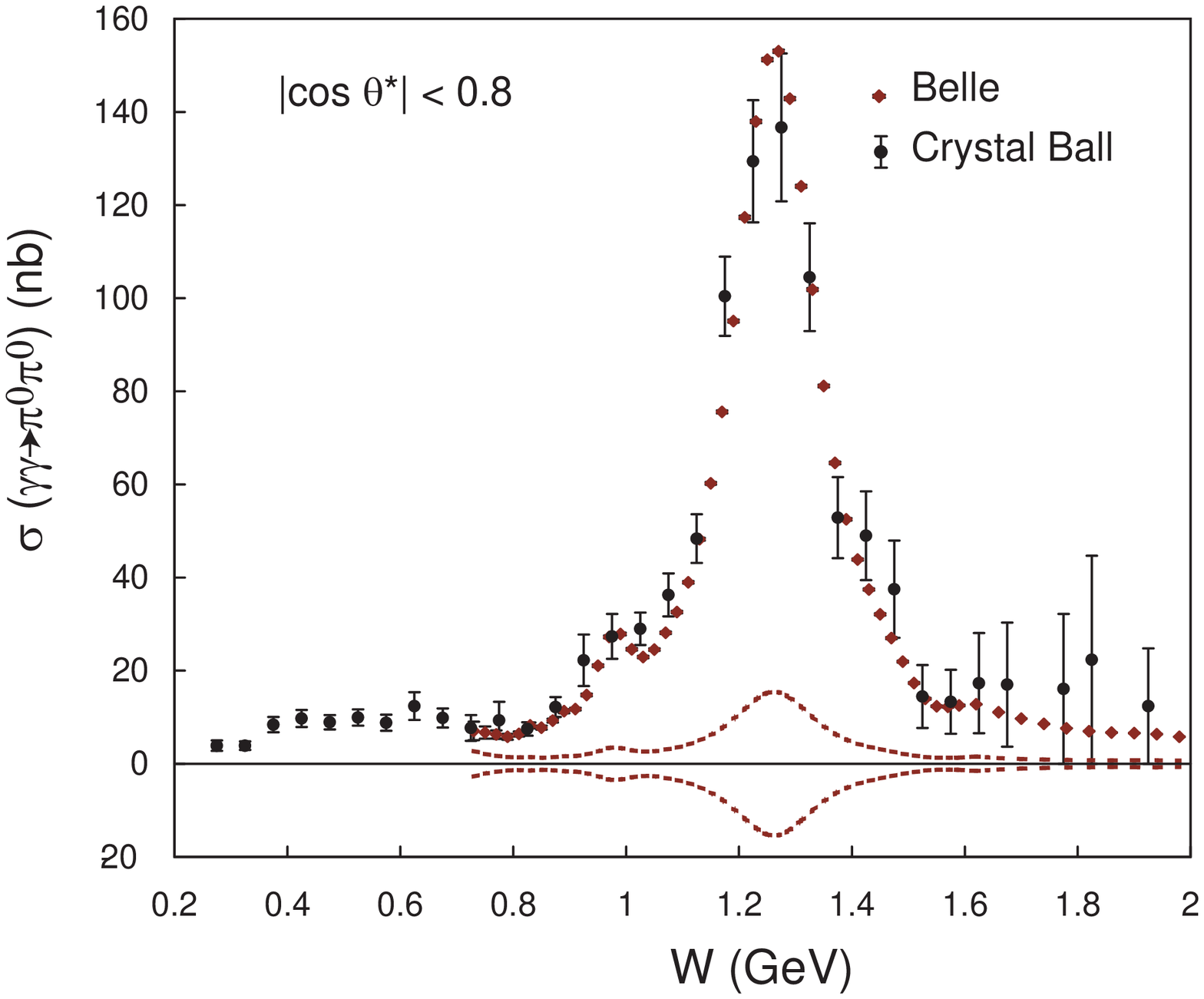}
   %\caption{\label{PRD79_fig2} 
  \end{minipage}
  \begin{minipage}{0.5\hsize}
   \includegraphics[width=7cm]{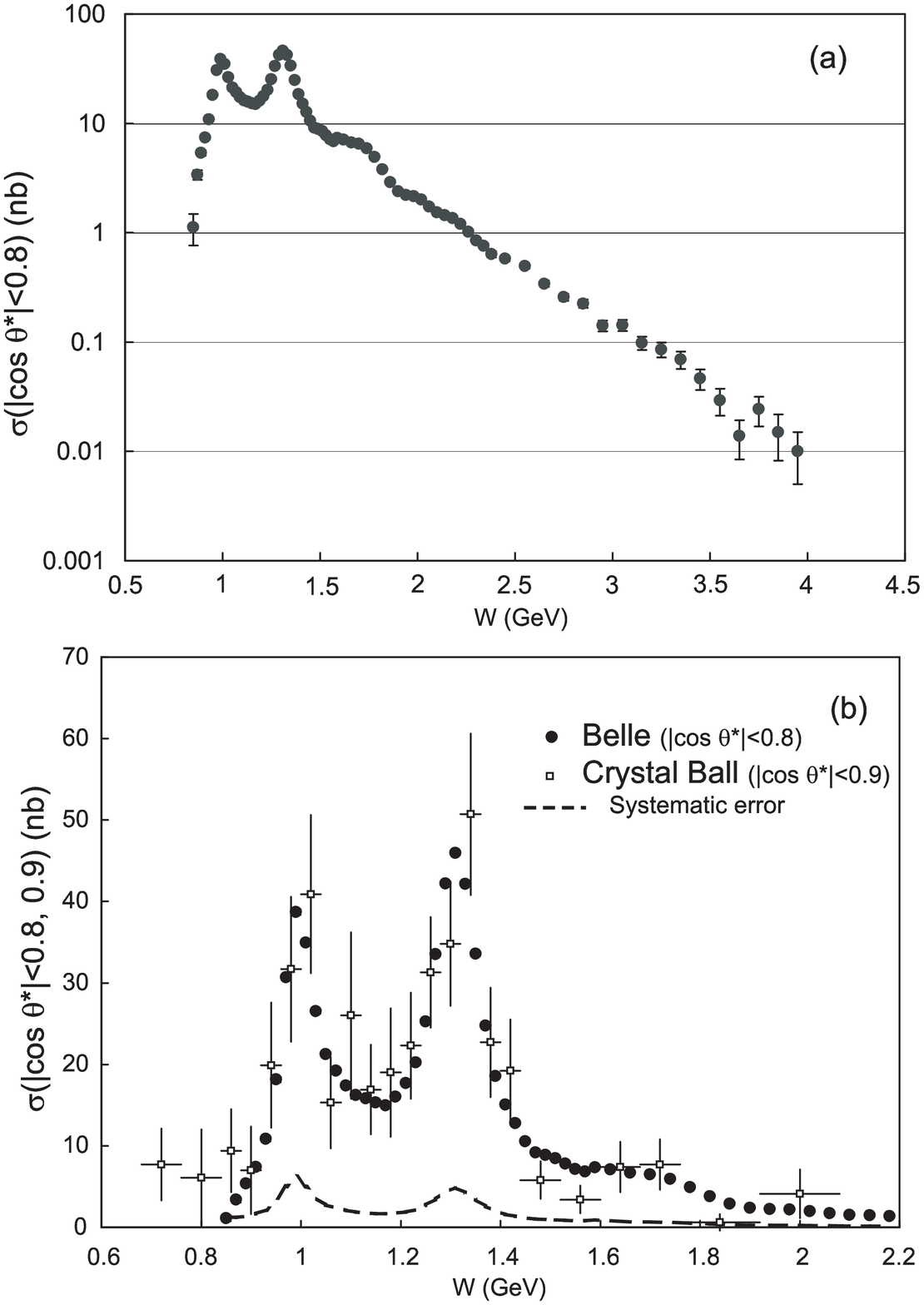}
  \end{minipage}
  \caption{%
  Left: (top)
  Cross section for $\gamma\gamma\to\pi^0\pi^0$
  ($|\cos\theta^*|<0.8$).
  Data points near 3.5~GeV are not shown because of
 uncertainty from the $\chi_{cJ}$ subtraction.
  (bottom)
  Comparison with the Crystal Ball measurement~\cite{cball2}.
  Dashed lines show the systematic errors for the Belle measurement.
  Right: 
  Cross section for $\gamma\gamma\to\eta\pi^0$ ($|\cos\theta^*|<0.8$)
  on (top) logarithmic and (bottom) linear scale compared with the Crystal Ball 
  measurement ($|\cos\theta^*|<0.9$)~\cite{cball}. The corrections for
  different $|\cos\theta^*|$ coverage are not made. The dashed 
  curve shows the size of the systematic error.}
  \label{cball} 
 \end{figure}
\end{center}
Figures~\ref{cball} (left) shows the total cross section
for $\gamma\gamma\to\pi^0\pi^0$
integrated over $|\cos\theta^*|<0.8$, 
where $\theta^*$ is a scattering angle of one of two mesons
with respect to the photon beam axis.
We observe clear peaks for the $f_0(980)$ near 0.98 GeV and the
$f_2(1270)$ near 1.25 GeV and find at least two more structures around
1.65 and 1.95 GeV.

We perform the partial wave analysis to the
$W$ (two-photon invariant mass) region
$0.8~{\rm GeV}<W<1.6~{\rm GeV}$
and
$1.7~{\rm GeV}<W<2.5~{\rm GeV}$ separately. 
In the lower energy region, in addition to $f_0(980)$, 
$f_2(1270)$ and
$f_2'(1575)$ we introduce a scalar meson $f_0(Y)$
to take into account a resonance-like structure around 1.2 GeV
in the $\hat{S}$ wave~\cite{pi0pi01}
which can be either the
$f_0(1370)$ or $f_0(1500)$ or a mixture of them.
In the higher energy region
we include $f_4(2050)$ and 
$f_2(1950)$.
Here we denote the latter as the ``$f_2(1950)$'' assuming that the $f_2(1950)$
is just an empirical parameterization representing any possible
resonances in this $W$ region~\cite{pdg}. 
The resulting parameters obtained by the fit
are listed in Table~\ref{tab1}. The $M(f_4(2050))$ and
``$M(f_2(1950))$'' flip and the widths are about two times larger than
their PDG values.
Although a more sophisticated model is necessary, our data clearly
require a $G$-wave component, and the unacceptably worse fit without the
$f_4(2050)$ strongly supports its finite two-photon coupling.

Figures~\ref{cball} (right) shows the cross section 
for $\gamma\gamma\to\eta\pi^0$ integrated over
$|\cos\theta^*|<0.8$,
on logarithmic and linear scales for partial $W$ regions.
The data points are in good agreement with those from Crystal
Ball~\cite{cball}.
We find three resonant structures: near 0.98~GeV ($a_0(980)$), 1.32~GeV
($a_2(1320)$) and 1.7~GeV (probably the $a_2(1700)$).
We focus on the region, $W<1.5~{\rm GeV}$, where $J>2$ waves can
be safely neglected, because a fit with $a_2(1320)$ parameters 
doesn't give a stable result.
Introducing $a_0(Y)$ to model the shoulder around 1.2~GeV in
the $\hat{S}^2$~\cite{etapi0} wave, 
we fit differential cross sections for the range
$0.90~{\rm GeV}<W<1.46~{\rm GeV}$.
The fit result is summarized in Table~\ref{tab1}.
\begin{center}
 \begin{table}
  \begin{tabular*}{150mm}{@{\extracolsep{\fill}}cccc}
   \hline
   &
   $\pi^0\pi^0$ (95/fb~\cite{pi0pi01}) &
   $\pi^0\pi^0$ (223/fb~\cite{pi0pi02}) &
   $\eta\pi^0$ (223/fb~\cite{etapi0}) \\
   \hline
   & $f_0(980)$ & $f_4(2050)$ & $a_0(980)$ \\
   Mass [MeV/$c^2$] & $982.2\pm 1.0$ & $1885^{+14}_{-13}$ &
	$982.3^{+0.6+3.1}_{-0.7-4.7}$ \\
   $\Gamma$ [MeV] & $285.5^{+17.2}_{-17.1}$ & $453\pm 20$ & $75.6\pm 1.6^{+17.4}_{-10.0}$ \\
   $\Gamma_{\gamma\gamma}(\pi^0\pi^0/\eta\pi^0)$ [eV] & & $7.7^{+1.2}_{-1.1}$ & $123^{+3+501}_{-2-43}$ \\
   \hline
   & $f_0(Y)$ & ``$f_2(1950)$'' & $a_0(Y)$ \\
   Mass [MeV/$c^2$] & $1469.7\pm 4.7$ & $2038^{+13}_{-11}$ & $1316.8^{+0.7+24.7}_{-1.0-4.6}$ \\
   $\Gamma$ [MeV] & $89.7^{+8.1}_{-6.6}$ & $441^{+27}_{-25}$ & $65.0^{+2.1+99.1}_{-5.4-32.6}$ \\
   $\Gamma_{\gamma\gamma}\mathcal{B}(\pi^0\pi^0/\eta\pi^{0})$ [eV] & $11.2^{+5.0}_{-4.0}$ & $54^{+23}_{-14}$ & $432\pm 6^{+1073}_{-256}$ \\
   \hline
  \end{tabular*}%
  \caption{Fit results for the light resonances.}
 \label{tab1} 
 \end{table}
\end{center}
\subsection{Analysis of the higher-energy region}
The leading-order QCD calculation~\cite{bl,bc} predicts
$d\sigma(\pi^0\pi^0)/d\sigma(\pi^+\pi^-)\approx 0.07$
at $|\cos\theta^*|=0$, changing to $\approx 0.4$ at $|\cos\theta^*|=0.6$,
and
$d\sigma(\eta\pi^0)/d\sigma(\pi^0\pi^0)=0.46 (f_\eta/f_{\pi^0})^2$
where $f_\eta(f_{\pi^0})$ is the $\eta(\pi^0)$ form factor,
while
$d\sigma(\pi^0\pi^0)/d\sigma(\pi^+\pi^-)=0.05$
by the handbag model~\cite{dkv}.
We can evaluate these predictions at $W>2.4~{\rm GeV}$ where
the contribution from resonances is small.

The power-law $W^{-n}$ dependence parameter of the total cross section
for $\gamma\gamma\to\pi^0\pi^0$
($|\cos\theta^*|<0.8$)
is obtained to be $n=8.0\pm 0.5\pm 0.4$, and
the cross section ratio to $\sigma(\pi^+\pi^-)$
is found to
be $0.32\pm 0.03\pm 0.05$ for $3.1~{\rm GeV}<W<4.1~{\rm GeV}$.
For $\gamma\gamma\to\eta\pi^0$
($|cos\theta^*|<0.8$)
$n$ is obtained to be $10.5\pm 1.2\pm 0.5$.
The $n$ values are summarized in Table~\ref{nvalues}
together with those from other processes.
The ratio $d\sigma(\eta\pi^0)/d\sigma(\pi^0\pi^0)$ is consistent with
leading-order calculation of 0.46 if $f_\eta/f_{\pi^0}$ is 1.
\begin{table}
 \begin{center}
  \vspace{-3mm} \footnotesize
  \begin{tabular*}{170mm}{@{\extracolsep{\fill}}lclclcc}
   \hline
   Process & $n$ & $W$ range (GeV) & $|\cos\theta^*|$ range &
   Reference\\\hline
   $\eta\pi^0$ & $10.5\pm 1.2\pm 0.5$ & 3.1-4.1 & $<0.8$ &
		   \cite{etapi0}\\
   $\pi^0\pi^0$ & $8.0\pm 0.5\pm 0.4$ & 3.1-4.1 & $<0.8$ & \cite{pi0pi02}\\
   $\pi^0\pi^0$ & $6.9\pm 0.6\pm 0.7$ & 3.1-4.1 & $<0.6$ & \cite{pi0pi02}\\
   $\pi^+\pi^-$ & $7.9\pm 0.4\pm 1.5$ & 3.0-4.1 & $<0.6$ & \cite{kkpipi}\\
   $K^+K^-$ & $7.3\pm 0.3\pm 1.5$ & 3.0-4.1 & $<0.6$ & \cite{kkpipi}\\
   $K_S^0K_S^0$ & $10.5\pm 0.6\pm 0.5$ & 2.4-4.0 & $<0.6$ & \cite{ksks}\\
   \hline
  \end{tabular*}%
    \caption{%
  Power-low dependence parameters of the cross
  sections $\sigma\propto W^{-n}$ in various reactions.}
  \label{nvalues}  
 \end{center}
\end{table}
 \begin{center}
  \begin{table}
   \begin{tabular*}{170mm}{@{\extracolsep{\fill}}llllll}
    \hline
    Process & Mass (MeV/$c^2$) & Width (MeV) & events & significance & Reference\\
    \hline
    $X(3915)$ & $3915\pm 3\pm 2$ & $17\pm 10\pm 3$ & $49\pm 14\pm 4$ &
		    $7.7\sigma$ & \cite{X3915} \\
    $X(4350)$ & $4350.6^{+4.6}_{-5.1}\pm 0.7$ & $13.9^{+18}_{-9}\pm 4$ &
		$8.8^{+4.2}_{-3.2}$ & $3.2\sigma$ & \cite{X4350} \\
    \hline
   \end{tabular*}%
   \caption{Measured values for charmoniumlike states $X(3915)$ and
   $X(4350)$.}
   \label{Xvalues}
  \end{table}
 \end{center}
\section{$X(3915)$ and $X(4350)$}
\begin{figure}
 \begin{minipage}{0.5\hsize}
  \includegraphics[width=8cm]{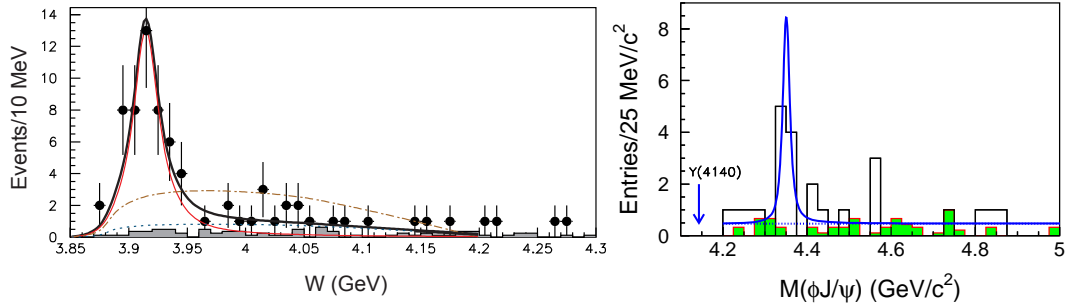}
 \end{minipage}
 \begin{minipage}{0.5\hsize}
  \includegraphics[width=6cm]{PRL104_fig3.epsi}
 \end{minipage}
 \caption{\label{figX} 
 Left: 
 $X(3915)$ in $\gamma\gamma\to\omega J/\psi$ 
 invariant mass distribution (dots with error bars) with 
 estimated background (shaded histgram)
 with curves from fit results~\cite{X3915}.
 Right: $Y(4350)$ in $\phi J/\psi$ invariant mass distribution
 (open histogram) with estimated background (shaded histogram).
 The solid curve is the best fit and the dashed curve 
 is the background~\cite{X4350}.}
\end{figure}

We search for charmoniumlike states in the 
$\gamma\gamma\to\omega J/\psi$~\cite{X3915} and 
$\phi J/\psi$~\cite{X4350} processes.
In the $\omega J/\psi$ invariant mass spectrum,
a peak is observed near $\omega J/\psi$ threshold (Figure~\ref{figX}).
Measured values for the peak (Table~\ref{Xvalues}) 
are consistent with those of $Y(3940)$~\cite{Y3940}
observed in the $\omega J/\psi$ final state,
and consistent with $Z(3930)$ seen in 
$\gamma\gamma\to D\bar{D}$~\cite{chic2'}, 
which is likely to be $\chi_{c2}^\prime$.

Motivated by a resonance-like peak named $Y(4140)$ found in 
$\phi J/\psi$ invariant mass spectrum by CDF~\cite{Y4140},
the $\gamma\gamma\to\phi J/\psi$ process is studied.
No $Y(4140)$ signal is observed. This disfavors the scenario in which
the $Y(4140)$ is a $D_s^{*+}D_s^{*-}$ molecule with $J^{PC}=0^{++}$
or $2^{++}$.
Instead, evidence of an unexpected new narrow structure, $X(4350)$
is found (Figure~\ref{figX} and Table~\ref{Xvalues}).
This is interpreted as a $c\bar{c}s\bar{s}$ tetraquark state with
$J^{PC}=2^{++}$~\cite{tetraquark}
or a $D_s^{*+}D_{s0}^{*-}$ molecular state~\cite{molecular} or
an excited $P$-wave charmonium state, 
$\chi_{c2}^{\prime\prime}$~\cite{chic2''}.

\section{Conclusion}
We have measured the diffential cross sections of the two-photon 
production of pure neutral final states, $\gamma\gamma\to\pi^0\pi^0$
and $\eta\pi^0$, using a high-statistics data sample collected with
the Belle detector at the KEKB accelerator. 
We perform the partial wave analyses to study
the light resonances.
In the higher energy region, QCD predictions are compared to the data.
The power-law dependence of the total cross section, 
$\sigma\propto W^{-n}$ and their ratios are presented.

We have observed a charmoniumlike peak $X(3915)$
in $\omega J/\psi$ invariant mass distribution and 
found evidence of a resonance-like structure $X(4350)$
in $\phi J/\psi$ mass spectrum, but no signal is 
observed at energy of $Y(4140)$.

 % \acknowledgments{We thank  $\cdots$.}
 
 \vspace{-2mm}
 \centerline{\rule{80mm}{0.1pt}}
 
\clearpage

\end{document}